\documentclass[showpacs,showkeys,superscriptaddress]{revtex4}
\usepackage{times}
\usepackage[]{graphicx}

\begin{document}

\keywords{atomic contacts, electrochemistry, gold, nanostructures, shell effects}
\pacs{73.23.-b, 73.40.Jn, 73.63.Rt, 68.65.-k}

\title[Formation of Atomic-Sized Contacts by Electrochemical Methods]{Formation of Atomic-Sized Contacts Controlled by Electrochemical Methods}
\date{October 6, 2006}

\author{M. Reyes Calvo}
\thanks{Corresponding
author: e-mail: {\sf reyes.calvo@ua.es}, Phone: +34\,965\,903\,540,
Fax: +34\,965\,909\,726} 
\affiliation{LT-NanoLab, Departamento de F\'\i{}sica Aplicada. Facultad de Ciencias (Fase II), Universidad de Alicante, Campus de S. Vicente del Raspeig, E-03690 Alicante, Spain.}

\author{Ancuta I. Mares}
\affiliation{Kamerlingh Onnes Laboratorium. University of Leiden, Leiden Institute of Physics, Postbus 9504, 2300 RA Leiden, The Netherlands.}

\author{Victor Climent} \affiliation{Instituto Universitario de Electroqu\'\i{}mica. Universidad de Alicante, Apartado 99, E-03080 Alicante, Spain.}

\author{Jan M. van Ruitenbeek}
\affiliation{Kamerlingh Onnes Laboratorium. University of Leiden, Leiden Institute of Physics, Postbus 9504, 2300 RA Leiden, The Netherlands.}

\author{Carlos Untiedt} \affiliation{LT-NanoLab, Departamento de F\'\i{}sica Aplicada. Facultad de Ciencias (Fase II), Universidad de Alicante, Campus de S. Vicente del Raspeig, E-03690 Alicante, Spain.}

\begin{abstract}
Electrochemical methods have recently become an interesting tool for fabricating and characterizing nanostructures at room temperature. Simplicity, low cost and reversibility are some of the advantages of this technique that allows to work at the nanoscale without requiring sophisticated instrumentation. In our experimental setup, we measure the conductance across a nanocontact fabricated either by dissolving a macroscopic gold wire or by depositing gold in between two separated gold electrodes. We have achieved a high level of control on the electrochemical fabrication of atomic-sized contacts in gold. The use of electrochemistry as a reproducible technique to prepare nanocontacts will open several possibilities that are not feasible with other methodologies. It involves, also, the possibility of reproducing experiments that today are made by more expensive, complicated or irreversible methods. As example, we show here a comparison of the results when looking for shell effects in gold nanocontacts with those obtained by other techniques.
\end{abstract}

\maketitle

\section{Introduction}

Electrochemistry is an old and wide field with an extraordinary number of applications. The introduction of electrochemical methods in the fabrication and study of nanostructures has attracted a lot of interest during the last years. A successful example of the application of electrochemistry to nanofabrication is the controlled deposition of metal nanowires into nanoporous templates \cite{taoreview}. This method has been also used in order to fabricate atomic-sized contacts with different metals and templates \cite{ninanowire,insitu}. However, this procedure is being mostly used as a tool for fabrication of nanowires with specific characteristics that are normally investigated by different techniques \cite{carlos,pirauxcobalt}.

The controllable fabrication of nanogaps and nanocontacts is crucial in the development of nanoscale electronic devices and electrochemistry might be the cheapest and easiest method to do it.
The first attempts to create nanocontacts using electrochemical methods were done in the late nineties \cite{snow,taofirst} where electrochemistry was used in combination with instruments like STM or AFM. It was Li et al. \cite{libogozitao} who showed that it is also possible to obtain contacts showing conductance quantization just dissolving a macroscopic wire, that is, obtaining atomic contacts with an extremely simple experimental setup. This result has been reproduced for different metals like Cu, Au, Ag, Pb,... \cite{linakato}. A slightly more complex method for electrochemical formation of atomic contacts in the case of zinc was reported by Nakabayashi et al. \cite{fractal}.

A point of great interest regarding electrochemistry is the reversibility of the reaction, allowing both to dissolve a wire or to create a contact from a gap. A controlled deposition of metal in between two electrodes reduces the distance of a macroscopic gap to the order of nanometers \cite{morpurgo,taoself} or even \aa{}ngstroms \cite{juanxiang}, which has the perfect size for attaching a molecule or to create easily an heterocontact of two different metals. \cite{heteropark}.

Electrochemical methods are not only a fabrication tool, they also provide a perfect environment to explore properties of the atomic contacts in solution \cite{taoraro}, or their behavior when different compounds are added to the medium. It is already possible to detect the presence of adsorbates \cite{moleculardetection} or other metals \cite{heavymetal} by measuring the conductance of small nanowires in solution. Moreover, interesting catalytic properties for monoatomic contacts in solution have been recently predicted \cite{teoria}.

Despite all the above considerations, the electrochemical fabrication of nanocontacts is still far from being an efficient technique at the level of others such as STM or MCBJ \cite{janreview}.
The experiments have a limited reproducibility and are usually time consuming and eventually uncontrollable. We have tried different methods of preparation for atomic-sized contacts of gold in solution. In this paper, we discuss different ideas proposed in the literature and our modifications in order to improve control or ease the fabrication.

Turning electrochemistry into a reproducible technique to prepare nanocontacts will open several possibilities that are not feasible with other instruments. It will involve, also, the possibility of reproducing experiments that today are made by more expensive, complicated or irreversible methods. As example, we show a comparison between the results obtained for shell structure in electrochemically fabricated gold nanocontacts with the ones obtained by MCBJ \cite{ancutashell}.

\section{Methods}

We have worked on the formation of gold atomic-contacts using electrochemical methods. Voltammetry was used to control the redox reaction for either dissolution or deposition of gold.
We have also tried different methods of preparation of samples and different experimental setups, following and modifying the ones already described in the literature.

\subsection{Samples and solution}

We have explored two different ideas for the sample preparation: a) starting from a macroscopic gold wire (0.1 mm of diameter) \cite{libogozitao}, it is dissolved until the desired size is reached or b) after sharpening two gold tips, they are placed in front of each other at a distance as short as possible (around 1 $\mu$m) and a contact is grown from this situation as it is described in section 2.3 below.

In both cases, we took special care to only expose the smallest possible area of our sample electrode. In this way, it is possible to keep the electrochemical current at very small values (less than $\mu$ A) in order to not interfere with the measurements of conductance. Different materials have been used for isolating the wires from the solution: tape, epoxy, etc. However, these materials are not as clean as it would be desirable and may lead to a fast contamination of the electrode surface. Moreover, our tests with epoxy resulted in the observation of contamination reflected in the presence of unexpected features in the voltammograms. Besides, after some hours of exposition, the solution leaked through the epoxy. We arrived to use polyethylene as best material to cover the samples. In the case when the experiment was started from a macroscopic gold wire, such wire was embedded in molten polyethylene which was cut afterwards in order to expose only around 100 $\mu$m of the wire to the solution. In the case of starting from the two opposed tips, the gold wires are sharpened by a common method to prepare STM tips \cite{sharp} and then covered by a drop of molten polyethylene, leaving only a small area of the sharpest part not covered. At the same time, this cover gives the wires enough rigidity to be mounted in the experimental setup. Polyethylene is quite inert and not porous, but after very long exposure to acid medium, it starts to suffer a slow degradation. However, no change was observable in the voltammograms, meaning that such degradation did not involve the addition of significant amount of contaminants to the solution.

After preparation, samples were cleaned with potassium permanganate solution. Once the samples are ready and clean, they can be immersed in the electrochemical cell. Cyanide is probably the most used electrolyte for gold dissolution \cite{morpurgo}, but it is also possible to dissolve gold using solutions containing sulfite, chloride or, even, iodine tincture \cite{iodine}. In this work we report results obtained with aqueous KCN and NaCl solutions. Cyanide is reported to lead to single-crystalline deposition of nanowires with the appropriate parameters \cite{cristornot}. This would be important in the case where single-crystallinity is relevant, but in the work reported here it is not critical whether the contact is polycrystalline or a single-crystal. In conclusion, the results with chloride and cyanide are very similar and chloride has the advantage of not being as toxic as cyanide. Moreover, opposite to cyanide, no spontaneous corrosion of gold in the presence of oxygen was observed in chloride solution.

Cyanide was used in a buffered solution, following the recipe given in \cite{morpurgo}. For chloride, 0.1 M H$_{2}$SO$_{4}$ solution was used as supporting electrolyte.
When we work with a macroscopic wire, we start to dissolve it in a 0.1 M NaCl solution (or KCN), lowering the concentration of the solution to 1mM NaCl+AuCl$_{3}$ (or KCN+K$_{2}$Au(CN)$_{3}$) when the conductance drops below 1000 G$_{0}$ (where G$_{0}$ is the quantum unit of conductance).
In case we start with the two tips we use only concentrations around 1 mM, avoiding in this setup the change of the solution.

\subsection{Voltammetry}

\begin{figure}[h]
\begin{minipage}[h]{.43\textwidth}
\includegraphics[width=\textwidth]{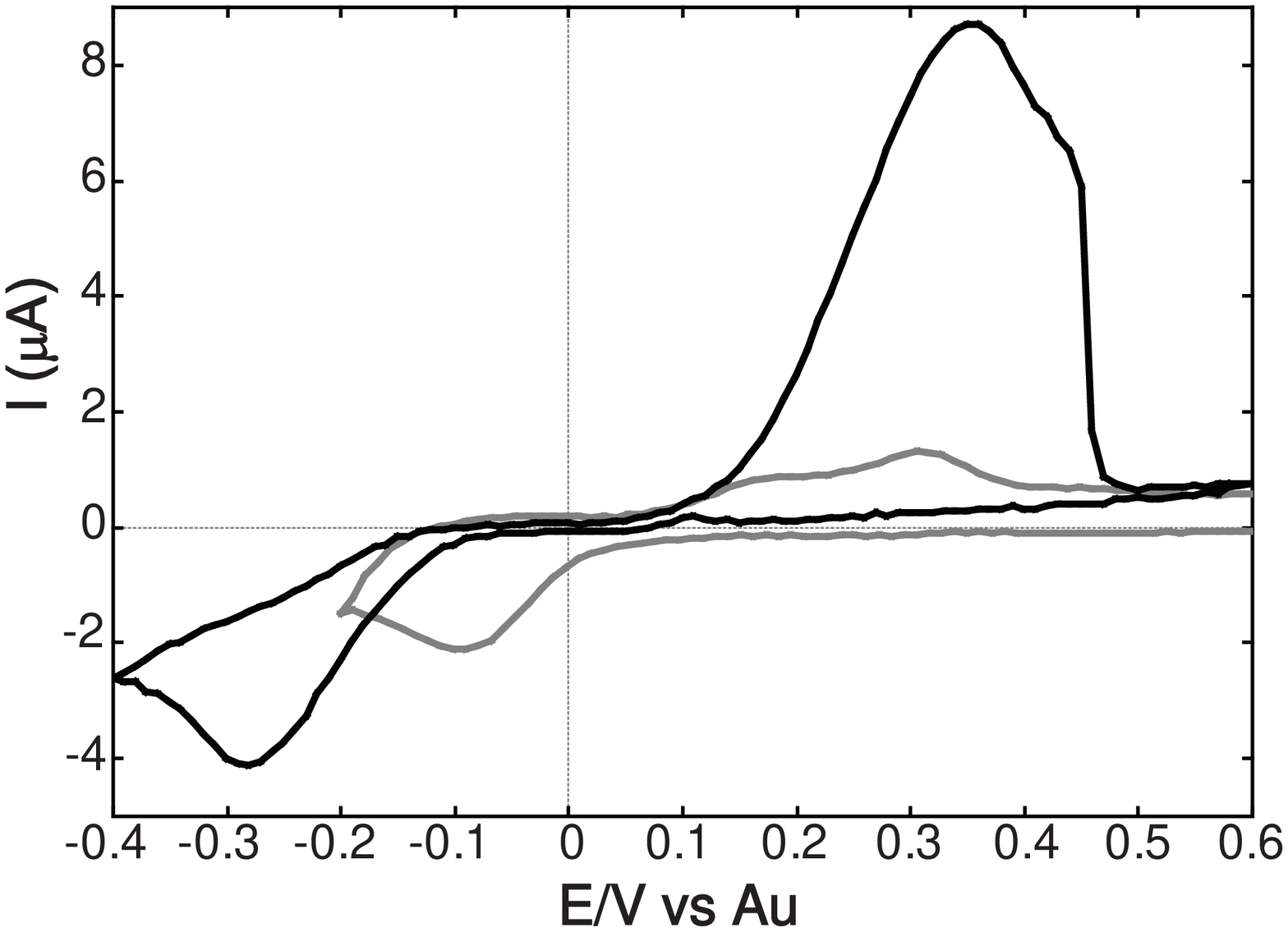}
\caption{A voltammogram for gold wire in a chloride solution 0.1M (black) and 1mM (grey) using a gold electrode as reference. (The grey line is multiplied by 10).}
\label{voltamograma}
\end{minipage}
\hfil
\begin{minipage}[h]{.52\textwidth}
\includegraphics [width=\textwidth] {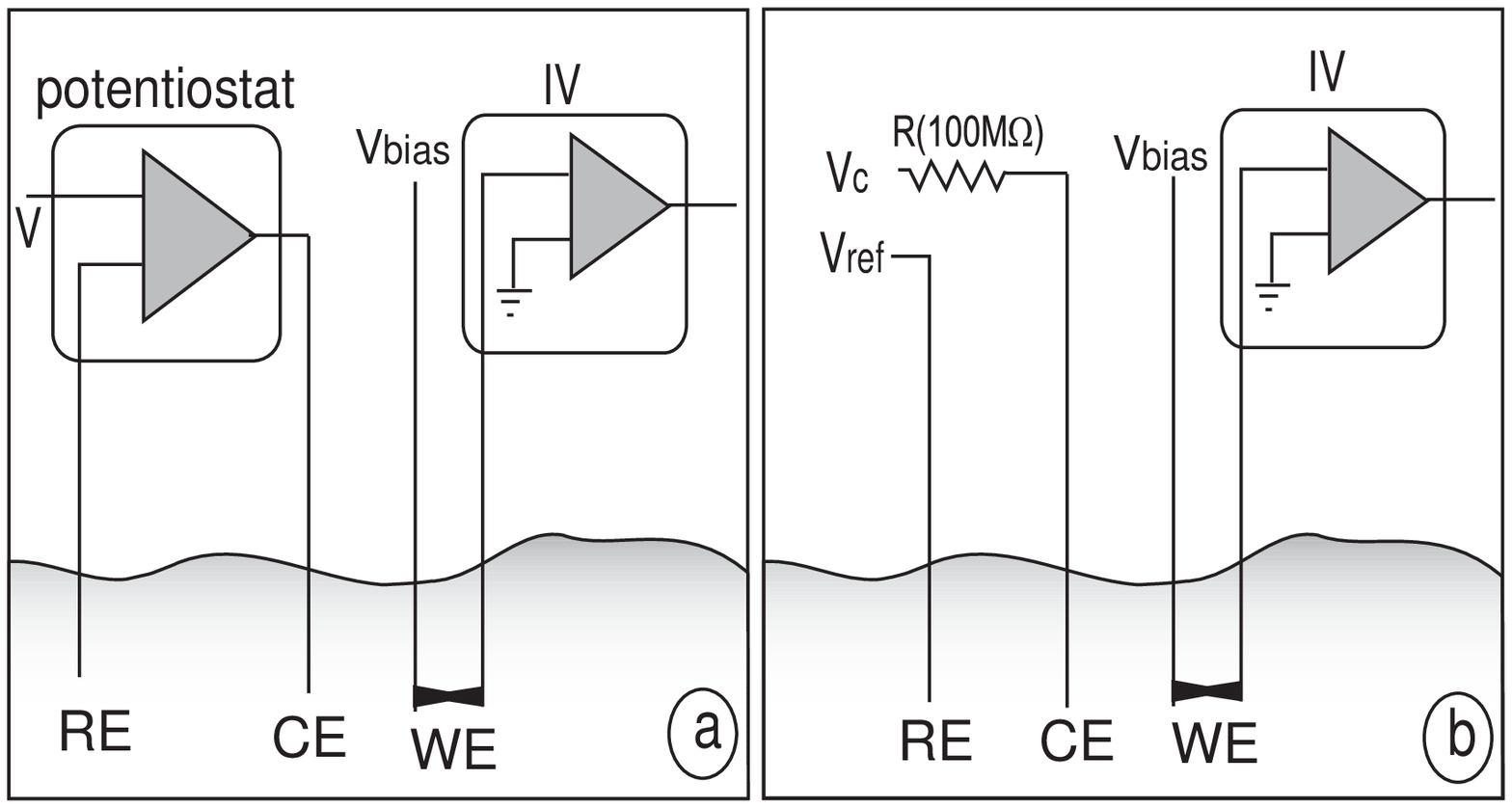}
\caption{ Sketches of the two described experimental setups.a) We use a potentiostat to dissolve or to deposit gold on the sample. b) Once contact is made, we apply a voltage Vc to the counter electrode. When the contact is broken we increase the bias to make a directional deposition between the electrodes}

\label{es}

\end{minipage}
\end{figure}

As shown in figure \ref{es}, the experimental setup is based on a three-electrodes electrochemical cell. With the depicted setup, it is possible to control the redox reaction in one of the electrodes (working electrode, WE) just by setting an input voltage (V) for the potentiostat. The potentiostat compares the input voltage with the one measured at the quasi-reference electrode (RE) and send the necessary current through the counter electrode (CE) in order to equalize these two potentials. In this case, the WE is the gold sample that is going to be used to form the atomic contacts and two extra gold wires (0.1 mm) are used as CE and RE.

Connecting the sample to a current-voltage converter it is possible to measure the electrochemical current flowing through the sample for a selected potential. A voltammogram is a record of the current measured between the CE and the WE when the voltage between the WE and the RE is linearly swept. The information provided by the voltammetry is of high value. On the one hand it helps us to choose the voltage ranges that are favorable for dissolving or depositing gold over the sample. Even in the case when the potentiostat is not used, as it will be explained in the section 2.3, we can still take advantage of the voltammetric information. On the other hand, contamination can usually be detected in the form of unexpected features in the voltammogram.

In figure \ref{voltamograma}, we show two examples of voltammograms obtained for different samples. The curves are recorded for samples in solution of 0.1M and 1mM of NaCl, respectively. The difference in the size of the features is in part due to the use of different concentrations but also because the curves are recorded for different samples with different exposed areas. Even in the case of 1mM, in which the voltammogram is more noisy, only the two expected peaks representing Au dissolution/deposition are clearly observed, showing that the sample and electrolyte are free of contaminants. The peak at positive voltage is related to the dissolution of gold in the WE and the one at negative potential corresponds to gold deposition. The abrupt fall after the dissolution peak in the voltammetric current is due to surface passivation of the electrode \cite{herrera}, therefore we avoided this range of high potentials.

\subsection{Experimental setup}

We combine two different methods in the experimental setup depicted in figure \ref{es}. The setup shown in \ref{es}(a) is the three-electrodes cell with the home-made potentiostat employed for the voltammetric tests, just with the addition of a small dc (or ac) bias voltage of 10 mV applied to one side of the sample while the other side is connected to the current voltage converter.  This setup is quite similar to the one used in ref.\cite{libogozitao,morpurgo}. In this way, measuring the current flowing through the wire due to the bias we obtain its conductance, as shown in the section 3 below.
With a simple software feedback, we can control the selected voltage in the potentiostat to automatically deposit gold when a contact is broken or to keep the contact at a given size or in between two selected values of conductance. 
The main disadvantage of this configuration is that the deposition to rebuild the contact is not localized, taking place all over the exposed area of the electrode, leading to an increase of its rugosity and a long time to reform the contact between the two leads of the WE.

The other option, shown in \ref{es} (b) is more similar to a two-electrodes cell. 
For dissolving a contact, we use a gold wire as CE and our sample as WE. A voltage is applied to this electrode (labeled as Vc in the figure) and a serial resistor of 100 Mohms can be optionally used to limit the flowing current (to nA), assuming the resistance of the solution is always lower and therefore not limiting. A bias voltage of 10mV is applied over the two sides of the contact to measure its conductance.
When the contact is broken, one side of the sample will act as CE and the other one as WE, \cite{taoself} and current will flow between them driven by the bias voltage. This can be increased to values between 0.2 and 1 V in order to favour the electrochemical reaction between the two leads of the sample.

In order to avoid high current densities once the contact is formed, the bias voltage is automatically lowered again to 10 mV. The software is programmed to automatically switch the adequate voltage for the different described situations.
When using this setup, the voltage on the RE is monitored. By comparing it to the voltammogram, we can adjust the applied voltage (Vc or Vbias) in order to control the speed of the reaction (slow deposition favours more robust formation of the contacts) and avoid to get into voltage regimes where the electrode passivation occurs.

With both experimental setups we can measure either ac or dc conductance. In the case of the ac method the measurement was performed with the use of a lockin amplifier (SR830 DSP). 
DC measurements have the inconvenience that the electrochemical current is measured superposed to the current across the contact. This current is large enough to be a problem only if the concentration of the solution or the size of the electrodes are large. Taking care about these two factors, dc measurements have some advantages over ac: a faster data acquisition and it avoids the ac conduction through the solution at high frequencies \cite{acconduction}.

In case we want the contact to stay at a desired size or to make cycles of dissolution and deposition without breaking the nanocontact, setup a) is of a great utility. Also, for the case in which we start the experiment dissolving a large wire, this setup is perfect for doing a fast dissolution of the macroscopic wire until it has an adequate size to start the experiments. However, the setup b) has the advantage of avoiding the problems derived from the less-localized deposition mentioned for the first one. As shown in \cite{taoself}, the deposition from one side of the sample to the opposite one is directional, leading to a faster and more efficient growing of the contact.

\section{Results and discussion}

As introduced in the previous section, we can record traces of conductance during the dissolution of a contact or the growing of a new one. The software allows us to do cycles of dissolution/deposition 
between two selected values of conductance or to keep the contact at a given conductance value.  We will discuss how controllable these processes are and show selected results.

\subsection{Traces and size control}

Using the simple feedback implemented in the data acquisition software, we can maintain the conductance of the contact at a given value. Conductance is directly related to the radius of the contact as discussed in \cite{cra}, so that by controlling the conductance the size of the contact is also controlled. As done in ref. \cite{libogozitao} by using an electronic circuit, the software feedback adjusts the potential to a deposition value if the contact becomes larger than the selected conductance or to a dissolution potential if the size of the contact drops below the desired value.

\begin{figure}[ht]
\begin{minipage}[h]{.5\textwidth}
\includegraphics[width=\textwidth]{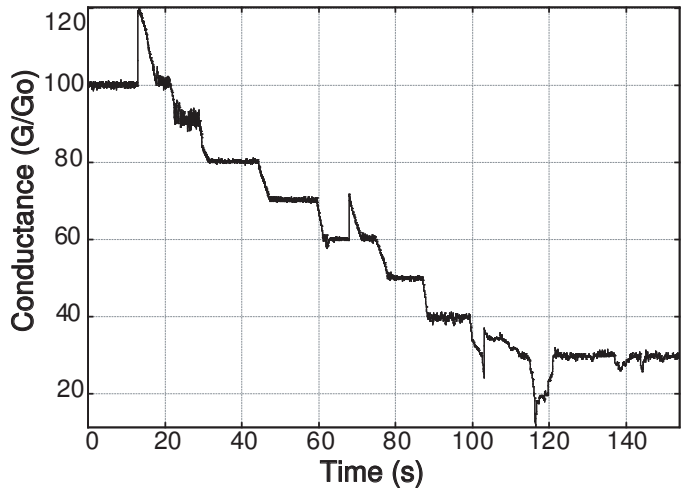}

\end{minipage}
\hfil
\begin{minipage}[h]{.4\textwidth}

\caption {A small contact is dissolved and the conductance is kept at different chosen values for long intervals of time. That is made with the help of a simple software feedback. Under 30 G$_{0}$ it becomes harder to control}
\end{minipage}
\label{controlling}
\end{figure}

In figure \ref{controlling} we show how a contact is dissolved with the help of the potentiostat and "artificially" stopped at different values of G$_{0}$ from 100 to 30. ($G_{0}$ is equal to $2e^{2}/h$ and denotes the quantum unit of conductance). Although sometimes the contact goes suddenly to larger values, after a short time it is corrected to the chosen value of conductance and stays there until we change the parameters. Below 30 G$_{0}$ it becomes more and more difficult to control and even when it is possible, most of the times the contact breaks spontaneously.

\begin{figure}[ht]
\begin{minipage}[h]{.45\textwidth}
\includegraphics[width=\textwidth]{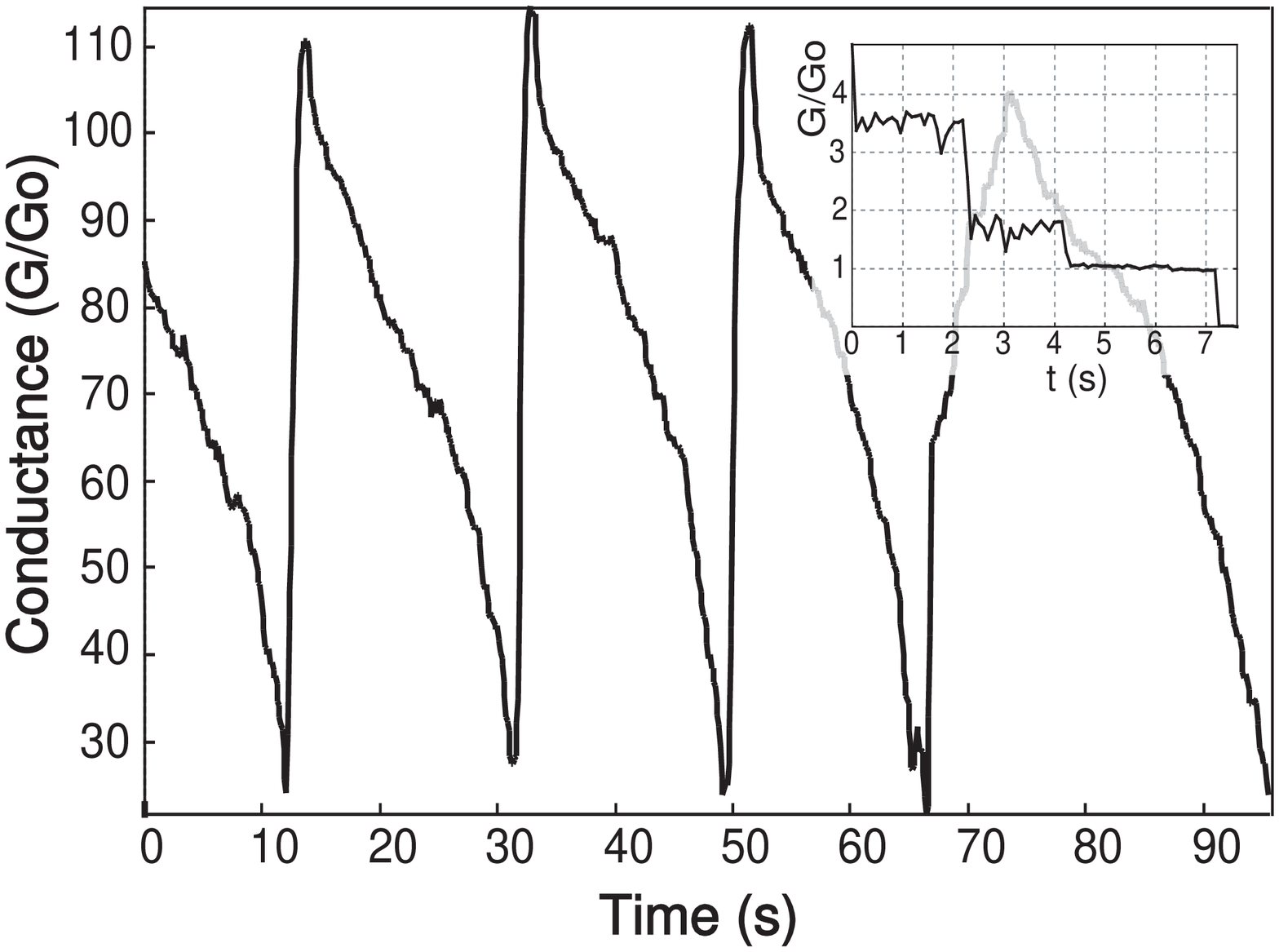}
\caption{Traces showing steps recorded during cycles of dissolution/deposition without breaking the wire. The inset shows an example of an atomic contact that remained stable for more than three seconds at 1 G$_{0}$}
\label{cycling}
\end{minipage}
\hfil
\begin{minipage}[h]{.45\textwidth}
\includegraphics[width=\textwidth]{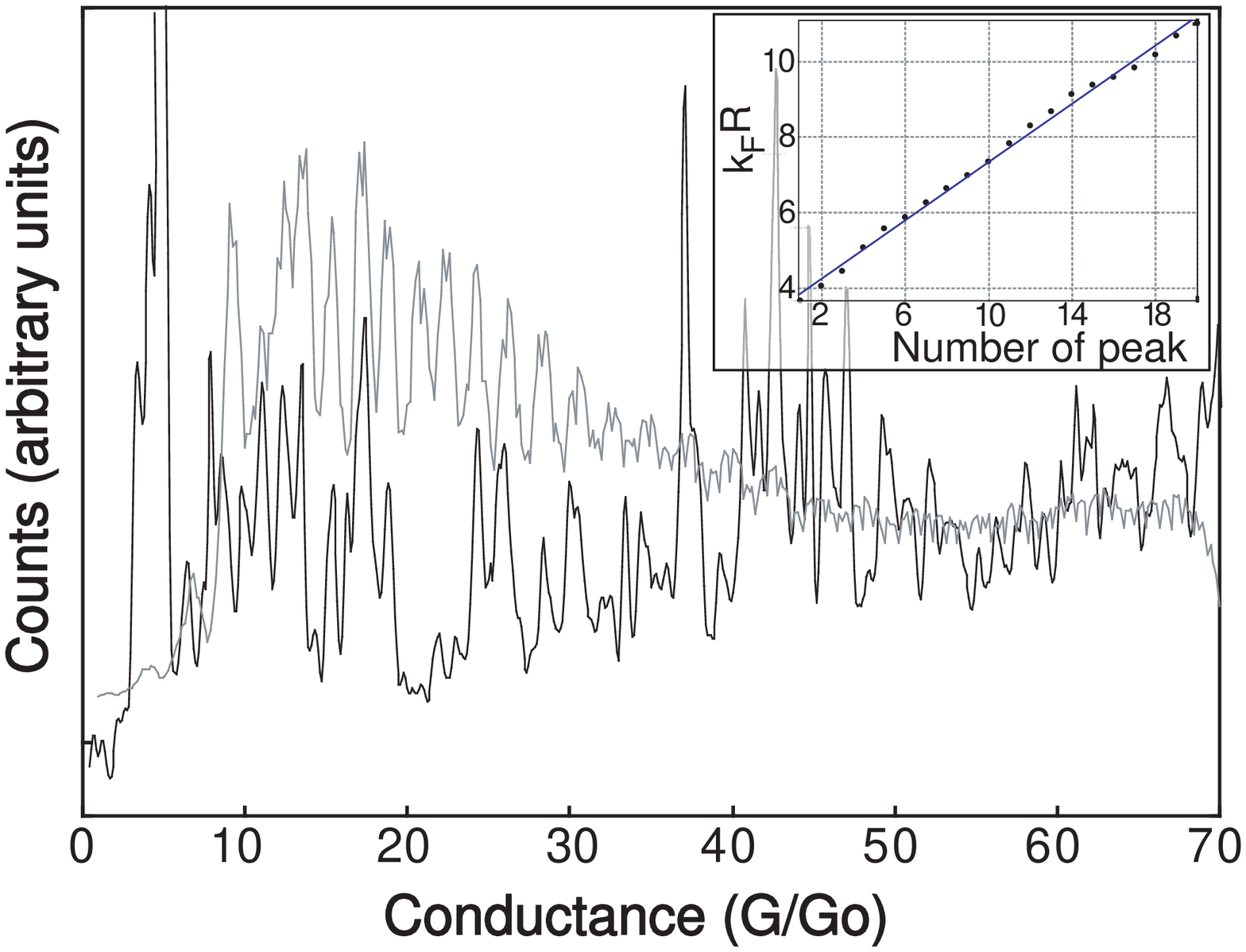}
\caption{Black: conductance histogram made from 145 scans from repeated dissolution/deposition of the wire. Grey:Histogram recorded for gold wire with an UHV-MCBJ. Inset: linear fit of radius versus peak number from data obtained in electrochemical environment.}
\label{shell}
\end{minipage}
\end{figure}

On the other hand, we are interested in the evolution of conductance when a small contact is dissolved or deposited just by applying a dissolution or a deposition voltage. For a fresh sample, it is observed in agreement with Li et al. \cite{libogozitao} that for the first cicles of dissolution and formation of the contact, traces do not show any conductance step neither conductance quantization at the last stages of contact breaking. We have observed that after some time working with the sample (keeping the conductance at a given value, then cycling it in between some values, etc...) this behavior changes, leading to the acquisition of traces with more structure. In figure  \ref{cycling}, we plot some of these cycles of dissolution/deposition on which it is possible to see clear steps. The contact is in this case allowed to grow until 100 G$_{0}$ and is dissolved until 20G$_{0}$. The reason for not going further is that when going to smaller values, the contact breaks.

In the cases when we let the contact evolve to rupture, it is possible to observe steps at low integer values of conductance in some of the traces. These "plateaus" are associated with conductance quantization. The monoatomic contact is surprisingly stable considering that the experiment is performed at room temperature. The inset of figure \ref{cycling} shows an example where the one-atom contact (shown as a "plateau" at 1$G_{0}$ conductance) was stable for several seconds. We have even observed monoatomic contacts lasting up to 10 seconds in some other traces.

\subsection{Indications of shell structure}

When dissolving a wire, it is expected that the time it takes to change the wire diameter will depend on the stability of such diameter. The existence of special stable diameters was first observed by Yanson et al. \cite{yanson} for sodium nanowires using a Mechanical Controllable Break Junction (MCBJ) at low temperatures. This phenomenom is due to the so-called shell-effect, that has been extensively studied for different materials \cite{ancutaother}. Specifically for gold, different techniques such as STM \cite{shellstm}  UHV-MCBJ \cite{ancutashell} and TEM \cite{tem} have been applied to the study of this phenomenom.

As we discussed in the previous section, when a fresh sample was dissolved we observed no steps of conductance, and therefore no indication of any shell-effect. However a conductance histogram constructed from a sufficiently large number of traces shows clear peaks. The peaks follow a certain periodicity. Since the conductance of a contact is related to its radius \cite{cra}, as introduced in section 3.1., the exhibited periodicity in the histogram could be a consequence of the existence of diameters with enhanced stability.

In figure \ref{shell}, we show a conductance histogram made from traces obtained with electrochemical methods (black line). It is possible to compare it with the histogram obtained from UHV-MCBJ data (grey line) and observe that, although noisier, the peak positions fit quite well. 
As described by Mares et al. \cite{ancutashell}, the radius of the contact is calculated from the conductance measurement. For shell effects, the peaks are expected to be equidistant when expressed as a function of the radius. In the inset of figure \ref{shell} the plot of $k_{F}R$ versus the peak number, (where k$_{F}$ is the Fermi wavevector and $R$ is the radius in $k_{F}^{-1}$ units), reveals a slope of $\Delta k_{F}R=0.4$ that agrees with the obtained result for atomic shell effect in gold in \cite{ancutashell}.

Despite the good agreement with previous data, to assert any further conclusion we should need to get better control or reproducibility of the experiments.
The breaking and formation processes are quite different to MCBJ ones and the existence of traces not showing any steps makes us to hesitate if shell effects should or should not be expected for electrochemically formed nanocontacts. A possible explanation may come from the configuration of the contact. An evolution of the structure of the contact could happen while working electrochemically with it. During the first dissolution cycles it could be formed more like a disordered neck not showing in this case any structure in the traces. After working a certain time, the contact can evolve to a short nanowire configuration as the one supposed for Break Junction contacts and show, in this case, shell effects. One can imagine this process as some kind of electrochemical annealing.

\section{Conclusions}

In this paper we have presented our experience reproducing the previous works in the field and introducing some new ideas about the experimental setup such as working with chloride, using polyethylene as an inert substance to cover the electrodes, starting from two sharp gold tips and the modification to the method exposed in the ref. \cite{taoself} in order to break the formed contact after the directional deposition. We would like also to point out the importance of cleanliness in this field and the usefulness of voltammetry as a test of surface contamination.

Although we have shown that it is possible to form and study atomic-sized contacts by electrochemical methods and have argued that electrochemistry is a technique with fabulous possibilities for nanofabrication (room temperature, low cost, reversibility, etc...), some improvements are still necessary to turn it into a systematic methodology to prepare and work with nanocontacts in solution.

As it has been already pointed out, low conductance contacts (under 20 G$_{0}$) become sometimes uncontrollable, not responding to the applied voltage as expected. 
In ref. \cite{teoria} it is claimed that electrochemical properties for monoatomic contacts are expected to be significantly different to the bulk ones. Moreover, the atomic contacts are supposed to react easier than bulk electrodes. Extrapolating this argument, we could think that once our contacts become small enough they are dissolved preferentially, leading to a fast and uncontrolled break of the contact. We should consider also the possibility of nanocontacts having a different electrochemical response than bulk to an applied voltage.

Shell effects are observable in histograms of conductance obtained by electrochemical methods, exhibiting similar features to the ones measured with a MCBJ. However, the results have still low reproducibility and data recording is time consuming. Our aim is to develop electrochemistry into a reproducible and systematic method to apply in nanofabrication.

\begin{acknowledgements}
This work is part of the research program of the ''Stichting FOM'' 
C.\,U.\ and R.\,C.\, acknowledge support of the Spanish MEC
through FIS2004-02356 and Generalitat Valenciana through ACOMP06/138.
C.\,U.\ and V.\ C.\ acknowledge financial support through the  "Ram\'on y Cajal" program of the Spanish MCyT.
We are grateful to Paramaconi Rodriguez for technical support and to M. Testa for his contribution in the first stages of this work.
\end{acknowledgements}


\begin{thebibliography}{10}
\bibitem{taoreview} H. He and N. J. Tao. Encyclopedia of Nanoscience and Nanotechnology X 1-18. American Scientific Publishers (2003)
\bibitem{ninanowire} F. Elhoussine, S. Matefi-Tempfli, A. Encinas and L. Piraux. Appl. Phys. Lett. {\bf81} 1681 (2002)
\bibitem{insitu} W. Wu, J. B. DiMaria, H. G. Yoo, S. Pan, L. J. Rothberg and Y. Zhang. Appl. Phys. Lett. {\bf84} 966 (2004)
\bibitem{carlos} T. G. Sorop, C. Untiedt, F. Luis, M. Kroll, M. Rasa and L. J. de Jongh. Phys. Rev. B {\bf67} 014402 (2003)
\bibitem{pirauxcobalt} M. Darques, L. Piraux, A. Encinas, P. Bayle-Guillemaud, A. Popa and U. Ebels. Appl. Phys. Lett. {\bf86} 072508 (2005)
\bibitem{snow} E. S. Snow, D. Park, and P. M. Campbell. Appl. Phys. Lett. {\bf69} 299 (1996)
\bibitem{taofirst} C. Z. Li and N. J. Tao  Appl. Phys. Lett. {\bf72} 894 (1997)
\bibitem{libogozitao} C. Z. Li, A. Bogozi, W. Huang and N. J. Tao.  Nanotechnology {\bf10} 221 (1999)
\bibitem{linakato} J. Li, T. Kanzaki, K. Murakoshi and Y. Nakato. Appl. Phys. Lett. {\bf81} 123 (2002)
\bibitem{fractal} S. Nakabayashi, H. Sakaguchi, R. Baba and E. Fukushima. Nanoletters {\bf1} 507 (2001)
\bibitem{morpurgo} A. F. Morpurgo, C. M. Marcus and D. B. Robinson. Appl. Phys. Lett. {\bf74} 2084 (1999)
\bibitem{taoself} S. Boussaad and N. J. Tao.  Appl. Phys. Lett. {\bf 80} 2398 (2002)
\bibitem{juanxiang} J. Xiang, B. Liu, B. Liu, B. Ren, Z. Q. Tian. Electrochem. Comm. {\bf8} 577 (2006)
\bibitem{heteropark} M. M. Deshmukh, A. L. Prieto, Q. Gu and H. Park.  Nanoletters {\bf3} 1383 (2003)
\bibitem{taoraro} C. Shu, C. Z. Li, H. X. He, A. Bogozi, J. S. Bunch and N. J. Tao. Phys. Rev. Lett. {\bf84} 5196 (2000)
\bibitem{moleculardetection} C. Z. Li, H. X. He, A. Bogozi, J. S. Bunch and N. J. Tao. Appl. Phys. Lett. {\bf76} 1333 (2000)
\bibitem{heavymetal} V. Rajagopalan, S. Boussaad and N. J. Tao.  Nanoletters {\bf3} 851 (2003)
\bibitem{teoria} E. P. M. Leiva, C. G. S\'anchez, P. Velez and W. Schmickler. Phys. Rev. Lett. B {\bf74} 035422 (2006)
\bibitem{janreview} N. Agrait, A. Levy-Yeyati and J.M. van Ruitenbeek. Physics Reports {\bf377} (2-3): 81-279 (2003)
\bibitem{ancutashell} A. I. Mares, A. F. Otte, L. G. Soukiassian, R. H. M. Smith and J. M. van Ruitenbeek, Phys. Rev. B. {\bf70} 073401 (2004)
\bibitem{sharp} L. A. Nagahara, T. Thundat, S. M. Lindsay: Rev. Sci. Instrum. {\bf60} 3127 (1989)
\bibitem{iodine} A. Umeno and K. Hirakawa. Appl. Phys. Lett. {\bf86} 143103 (2005)
\bibitem{herrera} J. Herrera-Gallego, C. E. Castellano, A. J. Calandra and A. J. Arvia. J. Electroanal. Chem. {\bf66} 207 (1975)
\bibitem{acconduction} A. Bardos, R. N. Zare and K. Markides. Chem. Phys. Lett. {\bf402} 274 (2005)
\bibitem{cristornot} J. Liu, J. L. Duan, M. E. Toimil-Molares, S. Karim, T. W. Cornelius, D. Dobrev, H. J. Yao, Y. M. Sun, M. D. Hou, D. Mo, Z. G. Wang and R. Neumann. Nanotechnology {\bf17} 1922 (2006)
\bibitem{cra} J. A. Torres, J. I. Pascual and J. J. Saez. Phys. Rev. B {\bf49} 16581 (1994)
\bibitem{yanson} A. I. Yanson, I. K. Yanson and J. M. van Ruitenbeek. Nature (London) {\bf 400}, 144 (1999)
\bibitem{ancutaother} A. I. Mares and J. M. van Ruitenbeek. Phys. Rev. B. {\bf 72}, 205402 (2005)
\bibitem{shellstm} E. Medina, M. Diaz, N. Leon, C. Guerrero, A. Hasmy, P. A. Serena and J. L. Costa-Kr\"amer. Phys. Rev. Lett. {\bf91} 026802 (2003).
\bibitem{tem} Y. Kondo and K. Takayanagi, Science {\bf289}, 606 (2000).
\end{thebibliography}
\end{document}